\documentclass[12pt, draftclsnofoot, onecolumn]{IEEEtran}
\usepackage[small]{caption}
\usepackage{graphicx,epsfig}
\usepackage{subfigure}
\usepackage{amsthm}
\usepackage{amsmath, amssymb, array}
\usepackage[noend]{algorithmic}
\usepackage[ruled,vlined, linesnumbered]{algorithm2e}
\usepackage{cite}      
\usepackage{verbatim} 
\usepackage{framed}
\usepackage[dvips, bookmarks, colorlinks=false, plainpages = false, ]{hyperref}
\theoremstyle{definition}

\theoremstyle{Theorem}

\DeclareMathOperator*{\argmax}{arg\,max}

\newcommand{\SNR}{\mbox{SNR}}
\newcommand{\SINR}{\mbox{SINR}}
\newcommand{\ICI}[2]{{\rm ICI}^{(#1,#2)}}
\newcommand{\SI}{\rm I_{self}}

\newcommand{\xref}[1]{Section \ref{#1}}

\usepackage{color}
\usepackage[dvipsnames]{xcolor}
\usepackage{colortbl}
\newcommand{\textred}[1]{\textcolor{red}{#1}}

\newcommand{\textgreen}[1]{{\color{Green}{#1}}}

\graphicspath{{./}}
\DeclareGraphicsExtensions{.jpg,.eps,.pnm}
\begin{document}

\title{Probabilistic Medium Access Control for Full-Duplex
Networks with Half-Duplex Clients}
\author{
Shih-Ying Chen, Ting-Feng Huang,
Kate Ching-Ju Lin,~\IEEEmembership{Member,~IEEE}, Y.-W. Peter
Hong,~\IEEEmembership{Member,~IEEE}, and Ashutosh
Sabharwal,~\IEEEmembership{Fellow,~IEEE}
\thanks{Kate Ching-Ju Lin is with Research Center for Information Technology Innovation,
	Academia Sinica, Taiwan  e-mail: katelin@citi.sinica.edu.tw.}
\thanks{Shih-Ying Chen and Ting-Feng Huang  are with Research Center for Information Technology
Innovation, Academia Sinica, Taiwan}
\thanks{Y.-W. Peter Hong is with Institute of Communications Engineering, National
Tsing Hua University, Taiwan}
\thanks {Ashutosh Sabharwal is with Department of Electrical and Computer
Engineering, Rice University, USA}
}

\maketitle
\IEEEpeerreviewmaketitle

\begin{abstract}
The feasibility of practical in-band full-duplex radios has recently
been demonstrated experimentally.  One way to leverage full-duplex in
a network setting is to enable {\em three-node full-duplex}, where a
full-duplex access point (AP) transmits data to one node yet
simultaneously receives data from another node.  Such three-node
full-duplex communication however introduces {\em inter-client
interference}, directly impacting the full-duplex gain.  It hence may
not always be beneficial to enable three-node full-duplex
transmissions.  In this paper, we present a distributed full-duplex
medium access control (MAC) protocol that allows an AP to adaptively
switch between full-duplex and half-duplex modes.  We formulate a
model that determines the probabilities of full-duplex and half-duplex
access so as to maximize the expected network throughput. A MAC
protocol is further proposed to enable the AP and clients to contend
for either full-duplex or half-duplex transmissions based on their
assigned probabilities in a distributed way.  Our evaluation shows
that, by combining the advantages of centralized probabilistic
scheduling  and distributed random access, our design improves the
overall throughput by 2.70$\times$ and 1.53$\times$, on average, as
compared to half-duplex 802.11 and greedy downlink-uplink client
pairing.
\end{abstract}

\section{Introduction}\label{Sec: Intro}


Recent works~\cite{sachin-mobicom10, sachin-mobicom11, sachin-sigcomm,
sachin-nsdi, ashu} have demonstrated the feasibility of in-band
full-duplex transmission using system level design and experiments.
Based on current experimental evidence, it appears that the first
potential adoption of full-duplex radios could be at AP serving
primarily half-duplex  mobile (thin) clients.
Thus, to fully exploit the capability of a full-duplex AP, it is
potentially beneficial for a network to leverage three-node
full-duplex~\cite{fd-sidechannel}, where downlink and uplink
transmissions can be to and from two different half-duplex clients.
The three-node full-duplex scenario however introduces the new
challenge of {\em inter-client interference} (ICI), which is caused by
the transmission of an uplink client to the  reception at a downlink
client. The presence of ICI implies that the doubling in spectral
efficiency due to full-duplex transmission may not be feasible, and
client pairing -- the choice of downlink and uplink clients -- has a
significant impact on the resulting network throughput.

Some methods have recently been proposed to select simultaneous
downlink-uplink clients for three-node full-duplex.  The first method
avoids ICI by picking nodes that are completely hidden from each
other~\cite{fd-mac}. This is however not only unfair, but also
underutilizes the full-duplex opportunities since pairing partially
interfering clients can possibly improve the network throughput.  The
second method, e.g., \cite{a-duplex}\cite{power-controlled-mac},
allows clients to randomly contend for uplink access, while selecting
the downlink client producing the maximal throughput according to the
ICI introduced by the winning uplink client. This method however can
only pair a proper downlink client for a given uplink client, which
could not fully extract the full-duplex gains by jointly paring the
downlink and uplink clients. The third method is weighted random
pairing \cite{ContraFlow}, where the downlink and uplink clients are
randomly chosen based on the reactively measured historical success
probabilities. Such an approach however only implicitly considers the
effect of ICI, and hence might perform even worse than half-duplex,
especially in the cases of medium to high ICI.
The last method optimizes client pairing by explicitly scheduling all
the transmission sequences with consideration of ICI and traffic
demands reported from all pairs of clients~\cite{Janus}.  However,
such centralized scheduling requires large signaling overhead, fails
to adapt to network dynamics, and, in general, has not gained traction
in IEEE~802.11 standardization.

In this paper, we present a {\em distributed} MAC protocol that
determines the best client pairings \emph{probabilistically} and
utilizes the full-duplex mode only when it achieves a higher network
throughput.  In particular, we adopt a probabilistic access mechanism,
which assigns to each full-duplex pairing and each half-duplex
transmission an \emph{access probability} so as to maximize the
expected network throughput while maintaining fairness.  The
high-level intuition of our probability assignment is to allow nodes
to best adapt between the half-duplex mode and the full-duplex mode
based on inter-client interference among clients.  The assigned access
probabilities can then be used as guidelines for both the AP and
clients to perform CSMA/CA-based contention and opportunistically pair
full-duplex transmissions in a distributed way.  Thus, our design
retains the desirable properties of 802.11's random access, while
adaptively exploiting the full-duplex opportunities.

A practical challenge needs to be solved is that the assigned access
probability should be no higher than the traffic demand of a client.
Otherwise, if we allocate more transmission opportunities than the
demand to a client, the spectrum resources will be wasted, leading to
lower channel utilization and throughput degradation. Therefore, we
should assign access probability with careful consideration of
heterogeneous traffic demands of clients.  To address this concern, we
propose an {\em epoch-based} MAC protocol, which updates access
probabilities for every epoch according to the short-term average
traffic arrival rate of each client.  Within each epoch, the clients
and the AP then adopt {\em prioritized} contention to share
transmission opportunities based on the updated access probability. By
combining the benefits of both centralized probabilistic scheduling
and distributed random access, nodes efficiently exploit the
full-duplex opportunities to best serve their traffic demands and
maximize the overall network throughput.

We conduct extensive simulations to evaluate the performance of the
proposed probabilistic-based MAC protocol.  The simulation results
show that our design achieves 2.70$\times$ and 1.53$\times$ throughput
gains over 802.11's half-duplex contention and greedy client pairing,
respectively.  The more-than-double gains over half-duplex 802.11 come
from both the ability to utilize full-duplex opportunities and to
reduce the number of collisions and overhead in our prioritized
contention process.

The rest of this paper is organized as follows.  Section
\ref{sec:related} summarizes recent works on full-duplex designs.  We
define our probabilistic MAC protocol in Section~\ref{sec:model}. The
proposed access assignment scheme is given in
Section~\ref{sec:assign}, and the probabilistic-based contention
protocol is presented in Section~\ref{sec:protocol}.  We evaluate the
performance of our design in Section \ref{sec:results}, and, finally,
conclude this work in Section \ref{sec:conclusion}.
\section{Related Work}\label{sec:related}
Recent works \cite{off-the-shelf, fd-mesh, sachin-mobicom10,
sachin-mobicom11, sachin-sigcomm, sachin-nsdi, ashu} have designed and
implemented full-duplex radios that cancel self-interference using
different techniques, including beamforming, RF chain cancellation,
digital domain cancellation and circulators. With the exciting results
of these full-duplex radio implementations, several later studies then
either theoretically analyze the full-duplex gains or design the MAC
protocols for full-duplex communications.  The works can be classified
into two categories.
%
%
%
%
%

\vskip 0.05in
\noindent {\bf Bidirectional full-duplex:} Several papers have studied
the gain of enabling bidirectional full-duplex links, where both the
transmitter and receiver are equipped with full-duplex radios.
Theoretical works model how hardware linearity~\cite{dynamic-range}
and power allocation~\cite{dynamic-power} affect the achievable
throughput of bidirectional full-duplex. In~\cite{dist-fd-mac}, the
authors propose a distributed MAC protocol~\cite{dist-fd-mac} to
enable bidirectional full-duplex in 802.11 networks, and analyze its
bandwidth and energy efficiency.  The gains of using multiple antennas
to enable half-duplex multiplexing or full-duplex are characterized
in~\cite{mimo-fd}. An adaptation scheme is then proposed
in~\cite{midu} to make the best choice between MIMO and bidirectional
full-duplex transmissions.

\vskip 0.05in
\noindent {\bf Three-node full-duplex:}
More recently, the three-node full-duplex scenario has also been
considered where two half-duplex clients communicate simultaneously
with a full-duplex AP.  FD-MAC~\cite{fd-mac} modifies 802.11's CSMA/CA
so that hidden nodes can be selected to form three-node full-duplex,
whereas~\cite{distributed-mac} allows any node introducing limited ICI
to join full-duplex transmissions.  Both the above approaches however
favor only part of the clients.
Later works~\cite{ContraFlow}\cite{Janus} then further take fairness
into consideration.  ContraFlow~\cite{ContraFlow} uses historical
transmission success probability of each pair of clients to implicitly
infer the degree of their ICI.  The client pairings are then
determined by exploiting the tradeoff between fairness and success
probability.  Estimation based on statistics however is not accurate
enough because an error might be caused by many factors other than
ICI, e.g., improper bit-rate selection.
PoCMAC~\cite{power-controlled-mac} lets the AP and the uplink client
control their transmit power. It solves an optimization problem of
finding the transmit power of the AP and the uplink client so as to
maximize the minimum SINR of the uplink and downlink transmissions.
Moreover, it gives a downlink client more chances to access the medium
if it experiences lower ICI or has a strong receiving power from the
AP.  However, as the number of clients increases, the average
throughput degrades due to the increasing collision probability. In
contrast, the average throughput of our design increases as the number
of clients increases because we efficiently assign full-duplex
opportunities, while adopting probabilistic-based contention to avoid
collisions.  A-Duplex~\cite{a-duplex} uses the capture effect to
establish full-duplex transmissions. That is, it aligns two packets
from the AP and the uplink client at the downlink client properly such
that the downlink client can recover its intended packet in the
presence of the interfering uplink client.  It then proposes a user
selection mechanism to only pair clients if a downlink client receives
a much higher signal strength from the AP than that from the uplink
client.  Both works~\cite{power-controlled-mac}\cite{a-duplex} still
enable clients to contend for uplink randomly, while only selecting
downlink clients to improve the throughput gain.  In contract, our
design jointly considers both uplink and downlink to seek for more
full-duplex gains and maximize the average total throughput.

Janus~\cite{Janus} collects the numbers of buffered packets and ICI
from all the clients, and schedules batch half-duplex/full-duplex
transmissions in a time-division manner to minimize the completion
time of the buffered packets.  Centralized scheduling however not only
requires an expensive overhead, but can also fails to adapt to network
dynamics.  Our proposed protocol differs from the above MAC design in
that it assigns full-duplex/half-duplex transmission probabilities
based on ICI among clients, but leverages probabilistic-based
contention to realize thees assigned access probabilities in a
\emph{distributed} manner. In addition, different from our prior
work~\cite{globecom} which only considers backlogged traffic, this
work considers a more general scenario where clients might have
various traffic demands and, hence, the access probability should be
adjusted accordingly.

\section{Problem Definition}\label{sec:model}
In this section, we first define the access probability assignment
problem for a three-node full-duplex network.  We will describe the
formal model and our proposed probability assignment algorithm
in~\xref{sec:assign}, and then design a distributed random access
control protocol that realizes the assigned access probabilities
in~\xref{sec:protocol}.

\subsection{Three-Node Full-Duplex Networks}
We consider a three-node full-duplex network, where only the AP has
full-duplex capability, i.e., being able to transmit and receive
simultaneously in the same band, while all the clients are only
equipped with a half-duplex radio.  For ease of representation, we
give each node an index. We denote the index of the AP by $0$, and the
index set of clients is $\mathcal{C}=\{1,2,\dots,C\}$.  The AP can
operate in either the full-duplex or half-duplex mode. In the
full-duplex mode, the AP transmits a downlink stream to client $i$,
while receiving an uplink stream from another client $j$ at the same
time. Let $P_{0,i}$ and $P_{j,0}$ be the powers used for transmission
from the AP to client $i$ and from client $j$ to the AP, respectively.
Then, the SINR at downlink client $i$ is given by
\begin{eqnarray}
	\SINR^{(i,j)}_{d} =
	\frac{P_{0,i}|h_{0,i}|^2}{\sigma_i^2+\ICI{i}{j}}=
	\frac{P_{0,i}|h_{0,i}|^2}{\sigma_i^2+P_{j,0}|h_{j,i}|^2},
	\label{eq:SINR-down}
\end{eqnarray}
where the superscript $(i,j)$ represents the full-duplex pairing
between downlink client $i$ and uplink client $j$, $h_{j,i}$ is the
channel coefficient from client $j$ to client $i$, and $\sigma^2_i$ is
the noise variance at downlink client $i$. Here,  $\ICI{i}{j}$
represents the inter-client interference caused by client $j$ at
downlink client $i$, and can be measured by $P_{j,0}|h_{j,i}|^2$.
Similarly, the SINR of uplink client $j$ is given by
\begin{eqnarray}
	\SINR_{u}^{(i,j)}=\frac{P_{j,0}|h_{j,0}|^2}{\sigma_0^2+\SI}=\frac{P_{j,0}|h_{j,0}|^2}{\sigma_0^2+P_{0,i}|h_{0,0}|^2},
	\label{eq:SINR-up}
\end{eqnarray}
where $h_{0,0}$ is the self-interfering channel after suppression and
the self-interference $\SI$ can be expressed as $P_{0,i}|h_{0,0}|^2$.
Similarly, we let $(i,0)$ and $(0,j)$ denote the half-duplex downlink
transmission to client $i$ and half-duplex uplink transmission from
client $j$, respectively. The SINR can be expressed similarly as in
Eqs.~\eqref{eq:SINR-down} and \eqref{eq:SINR-up} by setting
$P_{0,0}=0$.

A no\"{i}ve solution to involving two clients in a full-duplex
transmission is to match each downlink client with the uplink client
that produces the maximal data rate. This simple method however could
favor only certain clients for uplink transmission. In addition, two
clients can only be paired together in the full-duplex mode if they
both have traffic (one uplink and one downlink) simultaneously. We
hence need to further consider the traffic demand of each client as
pairing full-duplex transmissions.  However, it is not only
computationally expensive but also impractical to schedule
transmissions on a per-packet basis.  To avoid this complexity, we
alternatively pair clients according to their {\em average traffic
demands}. Specifically, the AP monitors the average uplink/downlink
traffic arrival rate of each client for a historical period of time.
Let $\lambda^{(i)}_d$ and $\lambda^{(i)}_u$ denote the downlink and
uplink traffic arrival rate, respectively, of client $i$.  We then use
this statistical information to allocate either full-duplex or
half-duplex transmission opportunities to clients.  Since different
clients might have various traffic demands, we assign a proper access
probability to each pair of clients, i.e., a full-duplex pair of
clients, $(i,j)$, or a uplink/downlink half-duplex client,
$(0,i)$/$(i,0)$, with consideration of their traffic demands and SINR.

\subsection{Candidate Pairs of Clients}
In a three-node full-duplex network, some node pairs might have strong
inter-client interference such that the downlink client cannot
reliably recover its packets. To avoid this undesirable situation, we
first filter proper candidate node pairs before probability
assignment. For a pair of clients $(i,j)$, given the downlink and
uplink transmission power $P_{0,i}$ and $P_{j,0}$, respective, we can
derive the SINR of both clients based on Eqs.~\eqref{eq:SINR-down} and
\eqref{eq:SINR-up}, and use the SNR-based rate adaptation
algorithm~\cite{esnr} to select the optimal bit-rate $\gamma$, i.e.,
modulation and coding scheme, from a set of available bit-rates
$\mathcal{R}$.  Then, the effective throughput at downlink client $i$
and uplink client $j$ can be computed as
\begin{eqnarray}
	r_d^{(i,j)}=\max_{\gamma\in\mathcal{R}}
	\gamma\cdot\mbox{PDR}(\gamma,\SINR_{d}^{(i,j)})
\end{eqnarray}
and
\begin{eqnarray}
	r_u^{(i,j)}=\max_{\gamma\in\mathcal{R}} \gamma\cdot\mbox{PDR}(\gamma,\SINR_u^{(i,j)}),
\end{eqnarray}
respectively.\footnote{We use $r$ to represent the effective
throughput with consideration of loss and errors, while using $\gamma$
to represent an available bit-rate, i.e., modulation and coding
scheme.} Here, the packet delivery ratio $\mbox{PDR}({\gamma,\SINR})$
is a function of bit-rate $\gamma{\in}\mathcal{R}$ and the SINR at the
corresponding receiver.  We  also define $r^{(i,j)}\triangleq
r_d^{(i,j)}+r_u^{(i,j)}$ as the total throughput under the full-duplex
pairing of downlink client $i$ and uplink client $j$. Then, in the
half-duplex mode, we have $r^{(i,0)}=r_d^{(i,0)}$ and
$r^{(0,j)}=r_u^{(0,j)}$.

In the full-duplex mode, we are interested in pairing clients $i$ and
$j$ that yield non-negligible throughput (larger than $\epsilon$) in
both the uplink and downlink, i.e., the set of index pairs
\begin{eqnarray}
	\mathcal{P}_{\rm full}\triangleq\{(i,j): i,j\in\mathcal{C},~i\neq j, r_d^{(i,j)},r_u^{(i,j)}>\epsilon\};
\end{eqnarray}
and, in the half-duplex mode, we are interested in the set of index pairs
\begin{eqnarray}
	\mathcal{P}_{\rm half}\triangleq\{(i,j): i=0 \text{ or } j=0, r^{(i,j)}>\epsilon\}.
\end{eqnarray}
Then, all candidate pairs are collected as a set
\begin{eqnarray}
	\mathcal{P}\triangleq\mathcal{P}_{\rm full}\cup \mathcal{P}_{\rm half}.
\end{eqnarray}
We allow the AP to adaptively switch between full-duplex and
half-duplex by allocating a proper proportion of spectrum resources to
each candidate pair in $\mathcal{P}$.  In particular, our goal is to
assign each candidate pair $(i,j) \in \mathcal{P}$, including
full-duplex pairs and {\em virtual} half-duplex pair, a probability
$p^{(i,j)}$ to access the medium.  The objective of this access
probability assignment problem is to maximize the {\em expected} sum
rate of the whole network, while maintaining fairness among clients.
The formal problem formulation and the proposed probability assignment
algorithm will be given in the next section.
\section{Access Probability Assignment}\label{sec:assign}
In this section, we first describe the framework of our epoch-based
access control, and then formally define our system model and the
proposed solution.

\subsection{Epoch-based Access Probability Assignment}
An intuitive way to enable three-node full-duplex transmissions is to
select a pair of clients, one with a downlink packet and the other
with a uplink packet at the same time. Such deterministic {\em
per-packet client pairing}, however, requires a coordinator, e.g., the
AP, to exactly know the real-time traffic demands of each client, and
can hardly be realized in practice. To avoid this centralized
per-packet scheduling, we alternatively adopt an {\em epoch-based
assignment} framework that pairs clients in a probabilistic way
according to their historical average traffic demands, while still
allowing clients to contend for medium access in a distributed way,
i.e., without any coordination.

\begin{figure}
	\centering
	{
		\epsfig{width=5in,file=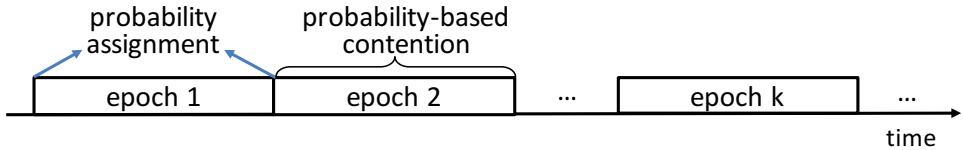}
		\caption{Epoch-based access probability assignment}\label{fig:epoch}
	}
\end{figure}

In particular, our system partitions the time slots into several time
intervals, called {\em epochs}, each of which has a duration $T$, as
shown in Fig.~\ref{fig:epoch}. The AP passively measures the average
downlink/uplink traffic demand of each client in each epoch.  In the
beginning of an epoch $e_t$, the AP uses the average traffic demands
measured from the previous epoch $e_{t-1}$ to find the proper access
probability $p^{(i,j)}$ for each candidate client pair $(i,j)$ in the
current epoch. The AP then notifies all the clients of their assigned
probability $p^{(i,j)}$ so that the clients can perform a
probabilistic-based contention in a distributed way according to their
assigned probability (see~\xref{sec:protocol}).

In theory, the access probability of a pair $p^{(i,j)}$ should be
defined as the probability that the pair occupies a unit of the time
slot.  However, note that the CSMA/CA mechanism of the standard 802.11
allows clients to share spectrum resources via frame-based
transmissions and guarantee packet fairness. That is, once a client
wins a transmission opportunity, it can send one data frame, no
matter how many bits are in the frame and what bit-rate is chosen. In
other words, each client has an equal probability to send its frames,
but the data frames from different clients might occupy different
channel time. To maintain the design philosophy of 802.11, i.e.,
packet fairness, we define the access probability of a candidate pair
as the proportion of transmission opportunities allocated to it.
Specifically, the access probability of a pair $p^{(i,j)}$ can be
expressed by
\begin{eqnarray}
	p^{(i,j)}=n^{(i,j)}/ \sum_{(i,j){\in}\mathcal{P}} n^{(i,j)},
	\label{eq:txop-to-prob}
\end{eqnarray}
where $n^{(i,j)}$ is the number of transmission opportunities obtained
by pair $(i,j)$ in an epoch. We can see from the equation that
assigning access probability is equivalent to allocating each pair a
proper number of transmission opportunities in an epoch. Therefore,
our probability assignment model aims at solving the variables
$n^{(i,j)}$ for all pairs $(i,j)\in \mathcal{P}$ and then finding the
corresponding probability $p^{(i,j)}$ accordingly.

\subsection{System Model and Algorithm}
Note that we have transformed the access probability assignment
problem into the problem of determining the number of transmission
opportunities in each epoch.  The allocation problem can be formulated
as follows:
\begin{subequations}
	\begin{align}
		& \text{\textbf{P}}_{\text{assign}}: &&\max
		\sum_{(i,j){\in}\mathcal{P}}\frac{n^{(i,j)}l^{(i,j)}}{T}
		\qquad \qquad \qquad & \label{eq:model-obj} \\
		&\text{subject to}	&&
		\sum_{j\in\{j:(i,j){\in}\mathcal{P}\}} n^{(i,j)}\geq	\eta^{(i)}_d, \forall
		i{\in}\mathcal{N}& \label{eq:model-1}\\
		& &&			\sum_{i\in\{i:(i,j){\in}\mathcal{P}\}}
		n^{(i,j)}\geq\eta^{(j)}_u, \forall
		j{\in}\mathcal{N} &\label{eq:model-2}\\
		&	&& 	\sum_{j\in\{j:(i,j){\in}\mathcal{P}\}} n^{(i,j)}\leq
		\lambda^{(i)}_d  T, \forall i{\in}\mathcal{N}& \label{eq:model-3}\\
		& && 	\sum_{i\in\{i:(i,j){\in}\mathcal{P}\}} n^{(i,j)} \leq
		\lambda^{(j)}_u  T, \forall j{\in}\mathcal{N}& \label{eq:model-4}\\
		& && \sum_{(i,j){\in}\mathcal{P}}n^{(i,j)}t^{(i,j)}\leq T &\label{eq:model-5}\\
		&\text{variables:} && n^{(i,j)}{\in} \mathbb{R}_{\geq 0},
		\forall (i,j){\in}\mathcal{P},&
	\end{align}
	\label{eq:model}
\end{subequations}
\noindent where $l^{(i,j)}$ represents the average total frame length
of pair $(i,j)$, i.e., $l^{(i)}_d + l^{(j)}_u$ (bits), and $t^{(i,j)}$
represents the average channel time occupied by each transmission of
pair $(i,j)$, which can be estimated by
\begin{eqnarray*}
	t^{(i,j)}=\max\left(\frac{l^{(i)}_d}{r_d^{(i,j)}},
	\frac{l^{(j)}_u}{r_u^{(i,j)}}\right).
\end{eqnarray*}
In our implementation, we use the average length of downlink/uplink
data frames as an estimate of $l^{(i)}_d$ and  $l^{(j)}_u$,
respectively.

\begin{algorithm}[t]
		\caption{Fairness Access Assignment}\label{algo:fairness_assignment}
		\SetKwInOut{Input}{Input}
		\Input {$\lambda^{(i)}_d$, $\lambda^{(i)}_u$, $\forall i{\in} \mathcal{N}$, $\bar{t}$, $T$}
		$d^{(i)}_d \leftarrow \lambda^{(i)}_d T, d^{(i)}_u \leftarrow \lambda^{(i)}_u T, \forall i{\in}\mathcal{N}$
		\linebreak
		\textsf{{\scriptsize // set of expected downlink and uplink traffic
		demands}}\\
		$\mathcal{D}=\{d^{(i)}_d: d^{(i)}_d > 0, i\in \mathcal{N}\} \cup \{d^{(i)}_u: d^{(i)}_u > 0, i\in \mathcal{N}\}$
		\linebreak
		\textsf{{\scriptsize // number of minimum shared transmission opportunities}}\\
		$\eta^{(i)}_d \leftarrow 0, \eta^{(i)}_u \leftarrow 0, \forall i{\in}\mathcal{N}$\\
		\While{$\mathcal{D}\neq \phi$} {
			$\eta_{left}\leftarrow \big(T-\sum_{i \in
			\mathcal{N}}(\eta^{(i)}_u + \eta^{(i)}_d) \bar{t}\big)\big/ \big( \bar{t}|\mathcal{D}|\big)$\\
			$\eta_{min} \leftarrow \min(\min_{d \in \mathcal{D}}d,
			\eta_{left})$\linebreak
			\textsf{{\scriptsize // terminate if the channel time of
			an epoch has be fully utilized}}\\
			\If {$\eta_{min} = 0$} {
				\textbf{break}\linebreak
			}
			\textsf{{\scriptsize // allocate $\eta_{min}$ to all the
			clients with demands}}\\
			\For{$i\in\mathcal{N}$}{
				\If {$d^{(i)}_d \ge \eta_{min}$} {
					$\eta^{(i)}_d \leftarrow  \eta^{(i)}_d + \eta_{min}$\\
					$d^{(i)}_d \leftarrow  d^{(i)}_d - \eta_{min}$\\
					\textbf{if } $d^{(i)}_d=0$ \textbf{ then }
					${D}\leftarrow
					\mathcal{D}\backslash\{d^{(i)}_d\}$\\
					}
				\If {$d^{(i)}_u \ge \eta_{min}$} {
					$\eta^{(i)}_u \leftarrow  \eta^{(i)}_u + \eta_{min}$\\
					$d^{(i)}_u \leftarrow  d^{(i)}_u - \eta_{min}$\\
					\textbf{if } $d^{(i)}_u=0$ \textbf{ then }
					${D}\leftarrow
					\mathcal{D}\backslash\{d^{(i)}_u\}$\\
					}
			}
		}
		{\bf return} $\eta^{(i)}_d, \eta^{(i)}_u$
\end{algorithm}

Eqs.~(\ref{eq:model-1}) and (\ref{eq:model-2}) represent the fairness
constraints. These constraints allow each client to obtain downlink
and uplink access opportunities no less than the minimum number of
guaranteed transmission opportunities $\eta_d^{(i)}$ and
$\eta_u^{(j)}$, respectively. We will discuss how to properly
configure the parameters of those minimum shares later.
Eqs.~(\ref{eq:model-3}) and (\ref{eq:model-4}) ensure that all the
scheduled traffic load does not exceed the arrived traffic demands at
the beginning of an epoch since we adopt epoch-based assignment.
Eq.~(\ref{eq:model-5}) ensures that the total allocated transmission
time should not exceed the epoch time $T$. Finally, the objective
is to maximize the expected total throughput of the whole network in
this epoch-based access control as shown in Eq.~(\ref{eq:model-obj}).

Note that the assignment problem \textbf{P}$_{\text{assign}}$ is a
linear programming (LP) problem, which can be easily solved by the
existing optimization solvers.  Once we solve the number of allocated
access opportunities $n^{(i,j)}$, we can then estimate the access
probability of each pair, $p^{(i,j)}$, based on
Eq.~\eqref{eq:txop-to-prob}.  The AP solves this probability
assignment problem at the beginning of every epoch, and notifies the
clients of the updated access probabilities $p^{(i,j)}$ in the beacon
frame.  Solving the probability assignment problem requires the AP to
know some information, e.g., the bit-rates $r^{(i,j)}_d$ and
$r^{(i,j)}_u$.  We will describe in
Section~\ref{sec:protocol-feedback} how the AP obtains these
information with an acceptable signaling overhead.

\vskip 0.05in
\noindent {\bf Minimum fair share:} The minimum allocated transmission
opportunities, $\eta^{(i)}_d$ and $\eta^{(j)}_u$, can be flexibly
configured by the system operator to realize their fairness policy. In
our implementation, we define that the fairness is achieved when each
client is guaranteed to obtain at least the same access opportunities
provided by traditional half-duplex CSMA/CA.  Specifically,
$\eta^{(i)}_d$ and $\eta^{(j)}_u$, respectively, are set to the access
opportunities for downlink client $i$ and uplink client $j$ in a fair
half-duplex protocol.  The AP adopts the max-min fairness to calculate
$\eta^{(i)}_d$ and $\eta^{(j)}_u$, as shown in
Algorithm~\ref{algo:fairness_assignment}.  Since these threshold
values are computed without considering channel variation, here, we
assume that the AP and the clients use the lowest bit-rate.  Then,
$\bar{t}$ represents the average half-duplex transmission time of all
the clients, i.e., the average frame length divided by the lowest
rate. The basic idea of the Algorithm~\ref{algo:fairness_assignment}
is that all the clients equally share the access opportunities in an
epoch unless the fair share exceeds one's traffic demand. In
particular, let $\mathcal{D}$ denote the set of unserved traffic
demands. We allocate the transmission opportunities to the demands in
$\mathcal{D}$ equally until all the demands are served, i.e.,
$\mathcal{D}=\phi$, or an epoch terminates, i.e., $\sum_{i\in
\mathcal{N}} (\eta^{(i)}_d + \eta^{(i)}_u) \bar{t} \ge T$.  After
obtaining the minimum shares, we can solve the probability assignment
problem using our algorithm summarized in
Algorithm~\ref{algo:probability_assignment}.

\begin{algorithm}[t]
		\caption{Access Probability Assignment}\label{algo:probability_assignment}
		Solve Algorithm~\ref{algo:fairness_assignment} to find the
			minimum shares $\eta^{(i)}_d$ and $\eta^{(i)}_u$\\
		Obtain the access opportunity $n^{(i,j)}$ by solving the assignment
				problem \textbf{P}$_{\text{assign}}$\\
		Transform $n^{(i,j)}$ to access probability
				$p^{(i,j)}$ based on Eq.~\eqref{eq:txop-to-prob}\\

		Announce $p^{(i,j)}$ to clients\\
\end{algorithm}

\vskip 0.05in
\noindent {\bf Feasibility:} Note that, with our defined minimum share
parameters $\eta^{(i)}_d$ and $\eta^{(j)}_u$, the assignment problem
\textbf{P}$_{\text{assign}}$ in Eq.~\ref{eq:model} is always feasible
because the half-duplex transmission assigned by
Algorithm~\ref{algo:fairness_assignment} meets all the constraints in
\textbf{P}$_{\text{assign}}$ and must be one of the feasible solutions
of \textbf{P}$_{\text{assign}}$.
\section{Probabilistic-based Full-Duplex MAC}\label{sec:protocol}
We next propose a probabilistic-based MAC protocol to realize the
access probabilities $p^{(i,j)}$ solved in
Algorithm~\ref{algo:probability_assignment}. The protocol addresses
practical issues including probabilistic access, bit-rate selection,
power control and information exchange to update probabilities.
\subsection{Probabilistic-based Contention} \label{sec:contention}
\begin{figure}
	\centering
	{
		\epsfig{width=5in,file=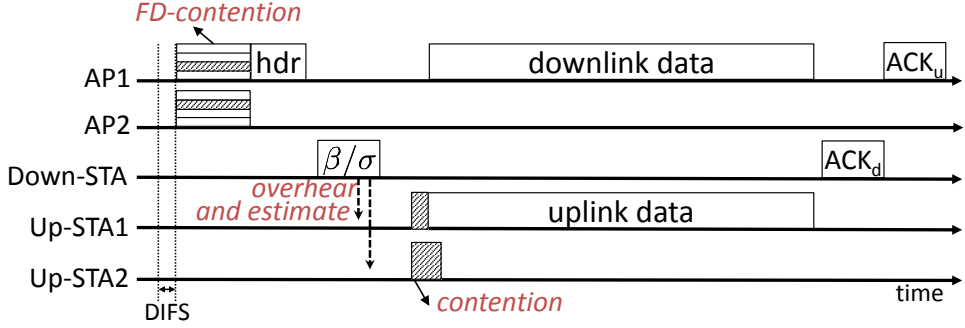}
		\caption{MAC protocol with interference-limited power control
	for down-up full-duplex networks}\label{fig:protocol}
	}
\end{figure}

To allow the clients to achieve the assigned access probabilities in a
distributed way, our protocol keeps the contention nature of 802.11,
and hence does not require the AP to coordinate the full-duplex
clients for every packet.  The AP determines to switch to half-duplex
or full-duplex based on the assigned probability $p^{(i,j)}$.  When
the AP enters the full-duplex mode, it adopts the so-called {\em
Down-Up} full-duplex\footnote{We focus on {\em Down-Up} full-duplex in
this work because it is easier to realize power control and bit-rate
selection in practice. However, our model actually can generally be
applied for {\em Up-Down} full-duplex, if those practical issues can
be addressed.}, where the AP first initiates its downlink stream,
before it allows the remaining clients to contend for uplink
transmission based on their assigned probabilities accordingly.  To
precede uplink contention with downlink transmission, we let multiple
APs leverage {\em frequency-domain contention}~\cite{FICA,Back2F} to
contend for downlink transmission first\footnote{To be able to coexist
with legacy 802.11 nodes, the APs can further contend for channel
access using traditional CSMA/CA before frequency-domain contention.},
as shown in Fig.~\ref{fig:protocol}.  When the channel becomes idle,
each full-duplex AP broadcasts a tone immediately after DIFS on a
randomly-selected sub-channel (formed by a few OFDM subcarriers), and,
at the same time, listens to the tones sent by neighboring APs. The
one that picks the smallest sub-channel, e.g., AP1 in
Fig.~\ref{fig:protocol}, wins the channel for downlink transmission.
The winning AP then randomly selects a downlink client $i$ with the
probability
\begin{eqnarray}
	p_d^{(i)}=\sum_{j\in\{j:(i,j)\in\mathcal{P}\}} p^{(i,j)}, \forall
	i\in\{0\}\cup\mathcal{N}.~\label{eq:down-prob}
\end{eqnarray}
If the index of the selected client $i$ is not `0', the AP sends the
PLCP and MAC header to downlink client $i$ after frequency-domain
contention; otherwise, if $i=0$, the AP keeps idle and the channel
time is reserved for half-duplex uplink transmission.  The remaining
clients can overhear the header of the downlink frame, and detect its
identity. Note that, idle channel implies that the downlink client is
`0'.  After sending the headers, the AP pauses its downlink
transmissions for the remaining clients to use traditional CSMA/CA to
contend for uplink transmission.  However, instead of using 802.11's
exponential backoff, the clients now contend for the uplink
transmission opportunity using {\em probability-based backoff}.
Specifically, each client $j$ sets its contention window to the
inverse of the following conditional access probability subject to the
maximum window constraint, i.e.,
\begin{eqnarray}
	CW_u^{(i,j)} = \min(\lceil 1/p_u^{(i,j)} \rceil, CW_{max}),\label{eq:cw}
\end{eqnarray}
where \begin{eqnarray}
	p_u^{(i,j)}=P(j\mbox{ wins uplink}|\mbox{AP sends
	to }i) =p^{(i,j)}/p_d^{(i)}~\label{eq:up-prob}
\end{eqnarray}
is the winning probability of client $j$, given the selected downlink
client $i$.  By doing this, a client with a higher access probability
has a smaller contention window, and can obtain an access opportunity
matching its assigned probability.  When the uplink client starts its
data frame, the AP then continues its downlink transmission
immediately, as shown in Fig.~\ref{fig:protocol}.

Note that we should reserve some channel time for half-duplex downlink
transmission when $p^{(i,0)} > 0$. If the clients always contend for
uplink transmission, the AP does not have any chance to enable its
downlink half-duplex transmission. To cope with this issue, we let the
AP perform uplink contention for the {\em virtual} uplink client `0',
which means no uplink transmission.  To do so, the AP computes
$CW_u^{(i,0)}$, randomly picks a backoff timer $w_u^{(i,0)}$ from
$[1,CW_u^{(i,0)}]$, and then announces the value of $w_u^{(i,0)}$ in
the downlink packet header. Each client $j$ gives up the uplink access
opportunity if its backoff timer $w_u^{(i,j)}$, randomly picked from
$[1,CW_u^{(i,j)}]$, is larger than $w_u^{(i,0)}$.

\subsection{Power Control and Bit-Rate Selection}\label{sec:power-control}
Since our design adopts {\em Down-Up} full-duplex, we should ensure
that the uplink client will not affect the decodability of the
existing downlink transmission. We address this problem by explicit
power control and bit-rate selection.  When the AP (or a client)
enters the half-duplex mode, it simply uses the maximum transmission
power $P$ to send downlink (or uplink) traffic and picks the bit-rate
based on its original SNR.  Without loss of generality, we assume that
the AP and each client~$i$ can passively learn its channel and
estimate its original SNR from beacons or ACKs.  On the other hand,
when the AP enters the full-duplex mode, it still uses full power to
transmit the downlink streams.  However, to ensure that the bit-rate
used by the downlink stream can be decoded properly, the uplink client
$j$ is instructed to control its transmission power and thereby the
ICI.  To simplify power control, for any downlink client $i$, the AP
picks the highest possible downlink bit-rate $\gamma_d^{(i)}$ that can
tolerate a fixed ICI $\delta$ (dB), i.e.,
\begin{eqnarray}
	\gamma_d^{(i,j)} = \argmax_{\gamma{\in}\mathcal{R}}
	\gamma\cdot\mbox{PDR}(\gamma,{\SNR}_d^{(i)} - \delta),
	\label{eq:downrate}
\end{eqnarray}
where $\SNR_d^{(i)}=P|h_{0,i}|^2/\sigma_i^2$ is the original SNR of
downlink client $i$ in the absence of interfering uplink transmission.
 By doing so, any uplink only needs to make sure that the
	residual ICI, after power control, should be no larger than
$\delta$ (dB).
Specifically, after uplink client~$j$ joins, the selected downlink
bit-rate $\gamma_d^{(i)}$ can only be decoded properly if
\begin{eqnarray}
	{\SINR}_d^{(i,j)} \ge {\SNR}_d^{(i)} - \delta,
	\label{eq:power_condition}
\end{eqnarray}
which can be rewritten as
\begin{eqnarray}
	&10\log_{10}(\frac{P|h_{0,i}|^2}{\sigma^2_i}) - 10\log_{10}(\frac{P|h_{0,i}|^2}{\sigma^2_i +
\ICI{i}{j} }) \le \delta\\
\Rightarrow & \frac{\ICI{i}{j}}{\sigma^2_i} = \frac{P_{j,0}^{(i,j)}|h_{j,i}|^2}{\sigma^2_i} \le
10^{{\delta}/10}-1,	\label{eq:power-control}
\end{eqnarray}
where $P^{(i,j)}_{j,0}$ is the transmission power of uplink client $j$
when $i$ is the concurrent downlink client. To ensure downlink client
$i$ is not harmed, client $j$, who wins the uplink transmission,
should adjust its transmission power $P^{(i,j)}_{j,0}$ to satisfy the
above constraint.

The nice part of such {\em interference-limited power control} is that
the AP can select the best downlink bit-rate $\gamma_d^{(i)}$ based on
the original downlink SNR without needing to know who will later win
the uplink contention.  However, since the channel might change
quickly, the challenge now is how to accurately estimate
$|h_{j,i}|^2/\sigma^2_i$ at the uplink client.  The estimation error
could increase ICI and consequently harm the decodability of the
downlink stream.  We hence enable per-packet inter-client channel
estimation, and ask the downlink client to announce its
$\beta/\sigma_i$ before uplink contention, as shown in
Fig.~\ref{fig:protocol}. The scalar $\beta$ is applied to bound the
amplitude of the announced signal within the hardware linearity range.
Each contending uplink client $j$ overhears
$y_j=h_{i,j}\beta/\sigma_i$ and can approximate
$|h_{j,i}|^2/\sigma^2_i$ by $(y_j/\beta)^2=|h_{i,j}|^2/\sigma^2_i$.
The client who wins the contention can use
Eq.~(\ref{eq:power-control}) to determine its power $P^{(i,j)}_{j,0}$
based on the estimated $|h_{j,i}|^2/\sigma^2_i$.  After power control,
the uplink client then selects its best bit-rate $\gamma_u^{(i,j)}$
based on the SINR, i.e.,
$P^{(i,j)}_{j,0}|h_{j,0}|^2/(\sigma^2_0+{\SI})$, as follows:
\begin{eqnarray}
	\gamma_u^{(i,j)} = \argmax_{\gamma{\in}\mathcal{R}}
	\gamma\cdot\mbox{PDR}(\gamma,
	P^{(i,j)}_{j,0}|h_{j,0}|^2/(\sigma^2_0+{\SI})).
	\label{eq:uprate}
\end{eqnarray}

\subsection{Information Exchange}\label{sec:protocol-feedback}
Solving the model \textbf{P}$_{\text{assign}}$ in Eq.~(\ref{eq:model})
requires the AP to know some information such as the bit-rates,
$r_u^{(i,j)}$ and $r_u^{(i,j)}$, and traffic arrival rates,
$\lambda_u^{(i)}$ and $\lambda_d^{(i)}$.  A simple way to estimate the
bit-rates is to collect the SINR information from clients and estimate
the bit-rates based on Eqs.~(\ref{eq:downrate},\ref{eq:uprate}).
However, to reduce the overhead of information feedback, we
alternatively let the AP estimate the downlink throughput
$r_d^{(i,j)}$ based on $({\SNR}_d^{(i)}-\delta)$ because the ICI has
been limited to at most $\delta$ via uplink power control.  We then
approximate $r_u^{(i,j)}$ by its best bit-rate $\gamma_u^{(i,j)}$,
which can be offline measured by client $j$ as mentioned in
Sec.~\ref{sec:power-control}.  Each client $j$ only needs to piggyback
in any uplink packet the index of the selected bit-rate
$\gamma_u^{(i,j)}$ for all $i{\in}\mathcal{N}$, which requires only
$\log_2{|\mathcal{R}|}(|\mathcal{C}|-1)$ bits.  Hence, the signaling
overhead can be reduced significantly, as compared to reporting the
information about SINR or the transmit power.  We will check in
Sec.~\ref{sec:results} how this approximation affects the performance.
On the other hand, the average traffic arrival rates,
$\lambda_u^{(i)}$ and $\lambda_d^{(i)}$, are offline measured at the
AP and clients. Each client $j$ also periodically piggybacks in any
uplink packet the measured average arrival rate of its uplink traffic
$\lambda^{(j)}_u$ to the AP.
\section{Performance Evaluation}\label{sec:results}
We conduct extensive simulations to evaluate the performance of our
probabilistic-based protocol.  A full-duplex AP and several
half-duplex clients are uniformly randomly distributed in an
$100m{\times}100m$ area. Each node has a maximum transmission power
15dBm and noise power -95dBm, respectively.  According
to~\cite{sachin-sigcomm}, we assume that each full-duplex AP can
cancel up to 110 dB of self-interference.  We use i.i.d.  complex
Gaussian channel with zero mean and unit variance and the log-distance
path loss model with the path loss exponent of 3.
The ICI threshold $\delta$ is set to 5 dB.  The set of available
bit-rates is $\{6, 9, 12, 18, 24, 36, 48, 54\}$ Mb/s, which are
supported in 802.11a, and the packet delivery ratio (PDR) of different
SNRs and bit-rates is measured using WARP platforms~\cite{warp} over a
2.4GHz band. The best bit-rate of each packet transmission is picked
based on the SNR and the measured PDR function. We use the Bernoulli
process with the time interval of 0.5 milliseconds as our traffic
arrival model.  Unless otherwise stated, we by default let each client
have backlogged (nearly unlimited) traffic demands, which can be
approximated by setting the traffic arrival rate to 2,000 frames/sec.
We will also evaluate the performance for limited traffic loads, and
the detailed configurations will be specified later. Each packet
contains 1,500 bytes.  Each simulation consists of 1,000 epochs, each
of which lasts for 100 milliseconds. We repeat each simulation for 5
random topologies, and report the average throughput.

We compare our design with the following baseline schemes.
\begin{list}{\labelitemi}{\leftmargin=9pt}
\setlength{\itemindent}{-4pt}
	\item {\em Oracle:} It is our design, except that the AP knows the
		exact information of the achievable throughput $r^{(i,j)}$,
		instead of the ones approximated by the best bit-rate, as
		mentioned in Section~\ref{sec:protocol-feedback}.
	 \item{\em MaxRate:} The AP randomly selects a downlink client,
		 and lets the client that achieves the maximal uplink bit-rate
		 join the full-duplex transmission without contention. This
		 scheme also applies our power control, and is used as an
		 upper bound for comparison.
	\item {\em Greedy:} The AP schedules deterministic pairing; that
		is, each client is only paired with one other client.
		Specifically, it greedily picks the pair $(i^*,j^*)$ from
		$\mathcal{C}_{\rm full}$ that produces the maximal throughput.
		All the other pairs $(i^*,j)$ and $(i,j^*)$, for all
		$i,j\in\mathcal{N}$, are then removed from $\mathcal{C}_{\rm
		full}$. The clients that cannot be paired then switch to the
		half-duplex mode. For each selected full-duplex pair, the
		downlink client is in charge of performing contention.
	\item {\em Random:} The AP randomly selects a downlink client, and
		the other clients use 802.11's CSMA/CA to contend for uplink
		transmission without considering ICI.
	\item {\em Half-duplex:} The traditional 802.11 MAC allows only
		half-duplex transmissions.
\end{list}
In Greedy, Random and Half-duplex, nodes use full power to transmit,
and select the proper bit-rate based on the SINR.

\begin{figure}[t]
\centering
{
\epsfig{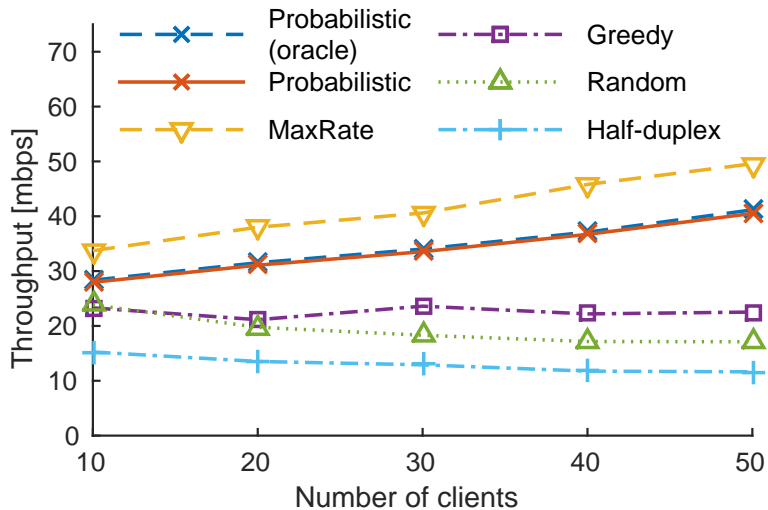} \\
\caption{Total throughput}
\label{fig:tput-backlogged-total}
}
\end{figure}

\begin{figure}[t]
\centering
{
\begin{tabular}{c}
\epsfig{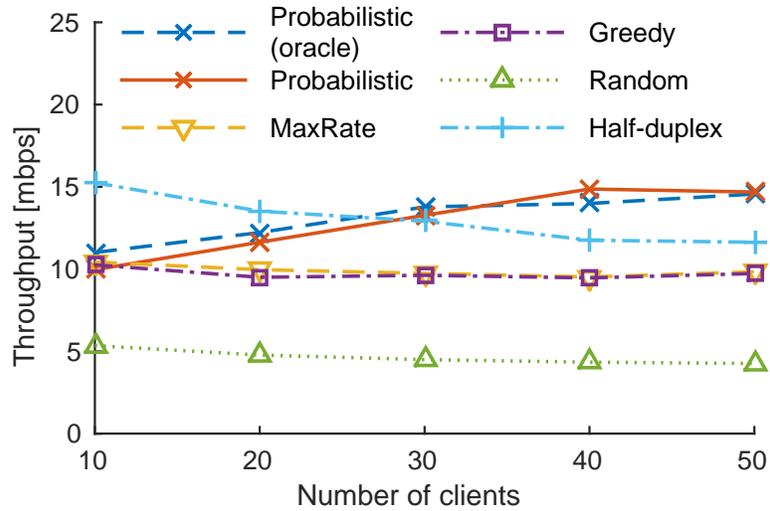} \\
(a) Downlink throughput
\\ \\
\epsfig{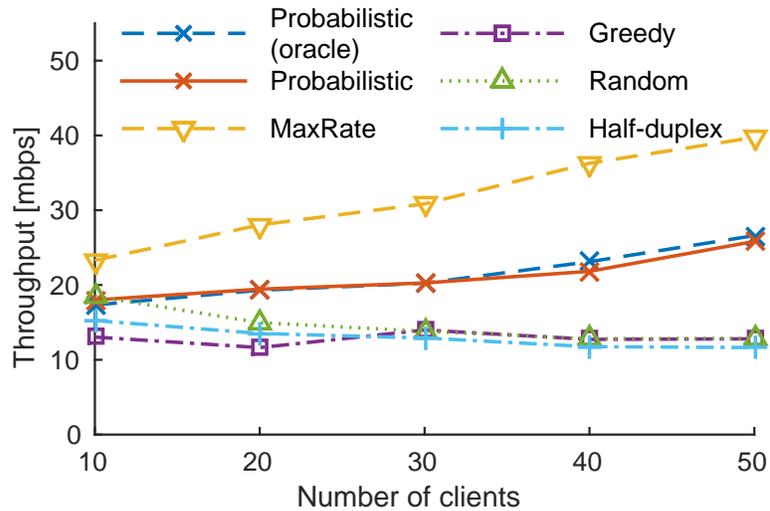}
\\
(b) Uplink throughput
\end{tabular}
\caption{Downlink/Uplink Throughput}
\label{fig:tput-backlogged}
}
\end{figure}

\subsection{Throughput Performance}
Fig.~\ref{fig:tput-backlogged-total} plots the average total throughput
(i.e., the sum of downlink and uplink throughput) of the comparison
schemes when the number of clients varies from 10 to 50.  The results
show that MaxRate, as expected, outperforms all the other schemes
because it favors some particular clients that are hidden to others.
We will however show later that some clients could starve in MaxRate
and cannot send any uplink traffic.  Greedy cannot efficiently utilize
the full-duplex opportunities because it runs in iterations, without
taking the overall throughput into account.  Consider a simple four
clients scenario. Say a possible pairing solution $\{(1,2), (3,4)\}$
produces the throughput 20 and 5 Mb/s, respectively, while another
pairing $\{(1,3), (2,4)\}$ produces the throughput 15 and 15 Mb/s,
respectively.  Greedy would output the first solution, which however
achieves a lower total throughput than the second one. Our
probabilistic-based paring performs better than greedy pairing and
random pairing, and improves the throughput by 1.53$\times$ and
2.70$\times$, on average, over Greedy and Half-duplex, respectively.
The gain increases when there exist more clients because more
candidate pairs are available for scheduling.  The gain over
half-duplex is sometimes more than doubling because {\em (i)} the
achievable throughput of different links are inherently different, and
{\em (ii)} more interestingly, our contention protocol prioritizes
clients by assigning different pairing probabilities, and hence
naturally decreases the number of collisions.  The results also show
that our design with approximated rate information performs similar to
that using full information (i.e., oracle). This means that the
quantized information does not affect efficiency of our design much,
but reduces the signal overhead significantly.

\begin{figure}[t]
\centering
\epsfig{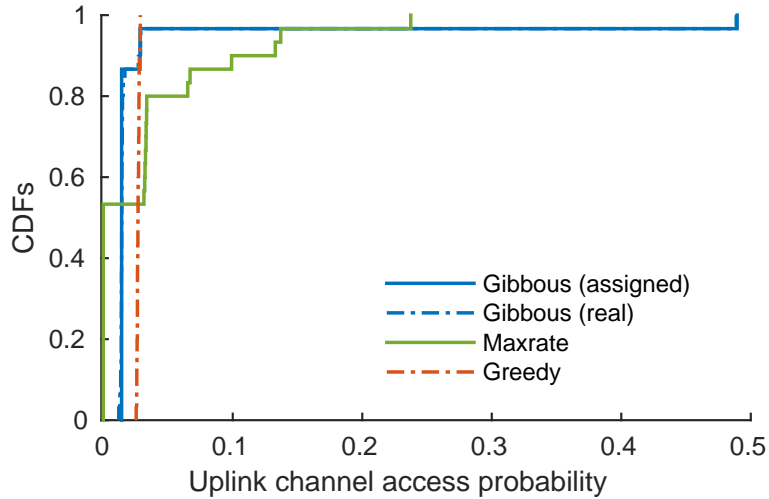} \\
\caption{CDFs of uplink transmission probability}
\label{fig:upaccess}
\end{figure}

To check where the gains come from, we further plot the average
downlink throughput and uplink throughput in
Figs.~\ref{fig:tput-backlogged}(a) and \ref{fig:tput-backlogged}(b),
respectively.  Random pairing significantly reduces the downlink
throughput because they might select a pair that introduces strong
ICI.  Greedy, MaxRate and our scheme all pair clients with
consideration of ICI, and hence do not affect the downlink throughput
much.  The results in Fig.~\ref{fig:tput-backlogged}(b) also answer
one's reasonable concern: Does uplink power control sacrifices the
uplink throughput?  We however observe that the clients achieve an
even higher uplink throughput than that in Half-duplex. This is
because, in our scheme, clients that introduce a smaller ICI have a
higher probability to participate in full-duplex transmissions.  These
clients also would not need to reduce their power levels
significantly.  More importantly, full-duplex communications allow
clients to better utilize the channel time for both downlink and
uplink access, and hence increase both downlink and uplink throughput.

MaxRate produces a much higher uplink throughput because it always
allocates the uplink transmission opportunities to certain clients
introducing negligible ICI.  To verify this point, we plot in
Fig.~\ref{fig:upaccess} the cumulative distribution functions (CDFs)
of the uplink access probability of clients, which is calculated by
the number of uplink transmissions occupied by one client divided by
the total number of uplink transmissions.  The figure shows that, in
MaxRate, 53\% of clients starve and cannot send any uplink traffic.
In contrast, our scheme ensures that all the clients at least obtain
their minimum share for uplink transmissions.

\begin{figure}[t]
\centering
{
\epsfig{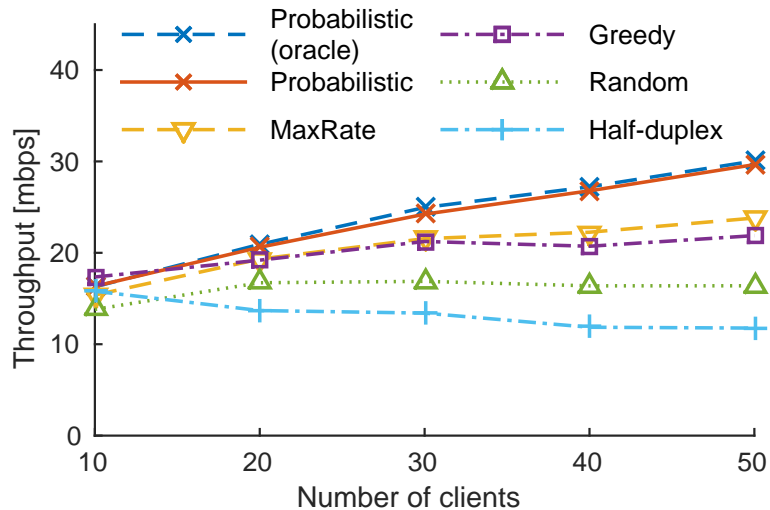}
\caption{Total throughput for heterogeneous traffic arrival rates}
\label{fig:random_traffic_arrival_rate}
}
\end{figure}

\subsection{Impact of Traffic Demands}
\label{sec:random-traffic}
We next check the impact of heterogeneous traffic demands on our
probabilistic-based MAC protocol. Specifically, in this simulation, we
assign each client a different traffic arrival rate, randomly picked
from 0 to 80 frames per second.  The choice of this demand range is
based on the observation that the network is mostly saturated when
each of 30 clients has a traffic demand of 16 frames per second.
Hence, the maximal demand, 80 frames per second, is about 5$\times$ of
the traffic arrival rate in the above saturated scenario.  We also
check a lower traffic arrival rate here to evaluate the performance of
our design when the network is not fully saturated.
Fig.~\ref{fig:random_traffic_arrival_rate} plots the average total
throughput when the number of clients varies from 10 to 50.  The
results show that our design again outperforms the comparison schemes
even when the network is not congested.  The MaxRate scheme now
performs worse than our scheme because, when the uplink traffic is
limited, instead of backlogged, the AP cannot always select those
uplink clients introducing negligible interference, but now needs to
allocate transmission opportunities to other pairs with stronger
inter-client interference. By contrast, our probability-based scheme
can flexibly fall back to half-duplex transmissions when no suitable
full-duplex pairs can be identified.  More importantly, the achievable
total throughput of our probability-based access control can scale
linearly with the increasing number of clients (and thereby the
increasing traffic load) when the network is not fully saturated.
However, the performance of the other comparison schemes converges
when the number of clients increases, since those greedy algorithms
might converge to a local optimum.

\begin{figure}[t]
\centering
\epsfig{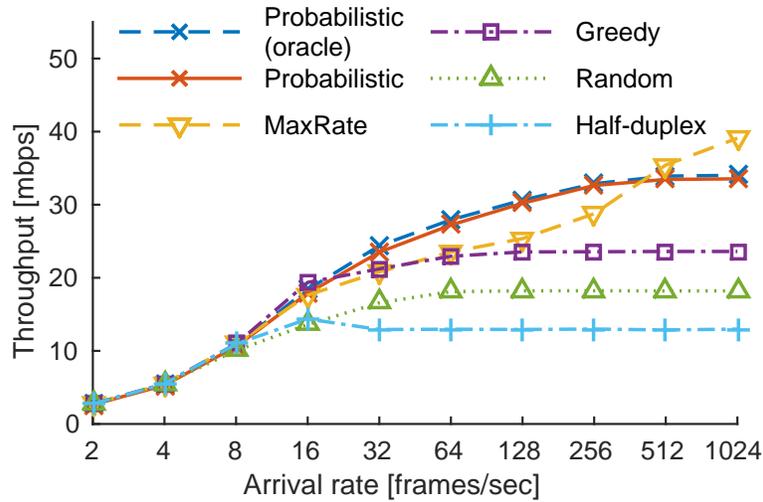}
\caption{Impact of traffic arrival rate}
\label{fig:traffic_arrival_rate}
\end{figure}

\begin{figure}[t]
\centering
{
\epsfig{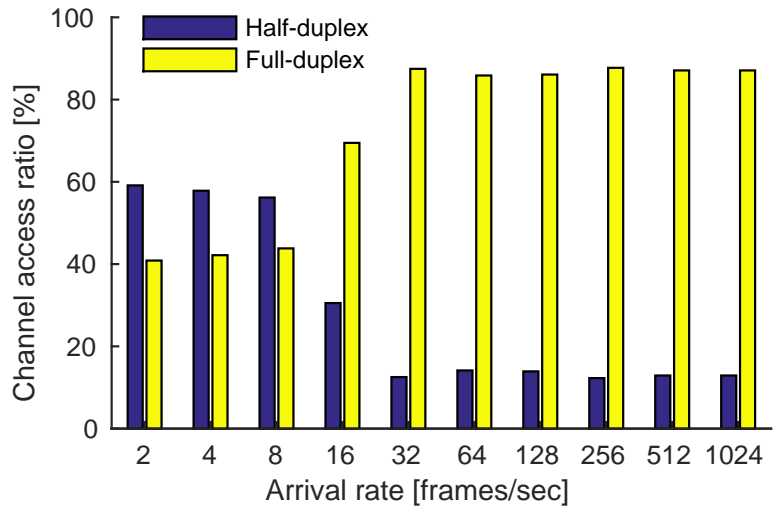}
\caption{Proportion of channel time occupied by full-duplex and
	half-duplex transmissions}
\label{fig:traffic_link_ratio}
}
\end{figure}

We further check the impact of the overall traffic load on the
throughput performance. In this simulation, we fix the number of
clients to 30, and assign all the clients a fixed traffic arrival
rate. The traffic arrival rate of each client is configured from 2 to
1024 frames/sec in different rounds of simulations. As a result, the
total traffic load increases when each client has an increasing
demand.  Then, the network is almost saturated when each of the 30
clients has a traffic demands of 16 frames per second. The average
total throughput for various traffic arrival rates is plotted in
Fig.~\ref{fig:traffic_arrival_rate}. The figure shows that the
throughput of half-duplex converges when the traffic demand increases
to 16 frames per second because the traffic load already saturates the
available spectrum resources.  The throughput of half-duplex then
degrades slightly when the traffic arrival rate exceeds 16 frames/sec
due to the increasing probability of collisions.  The throughput of
all the other comparison schemes, however, can grow and converge later
when clients have a larger demand because of the benefit of
full-duplex opportunities.

The results also show that the gap between our probabilistic-based
protocol and other greedy schemes increases when the traffic load
increases. This is because when clients have more traffic demands, an
efficient user pairing algorithm can better identify those suitable
pairs and extract more full-duplex gains. To verify this argument, we
further plot the proportion of channel time occupied by full-duplex
and half-duplex transmissions, respectively, in
Fig.~\ref{fig:traffic_link_ratio} for various traffic arrival rates.
The figure shows that our probability-based protocol can flexibly
switch between the half-duplex mode and the full-duplex mode according
to the traffic demands. Also, when traffic demands are large
sufficiently, in most of the channel time (around 90\%), we can
identify full-duplex pairs producing a rate higher than half-duplex
transmissions, and effectively utilize the full-duplex capability of
the AP.  Due to the same reason, our scheme performs better than
MaxRate when the clients have a lower traffic demand and, thereby,
those non-interfering pairs cannot always be matched as full-duplex
transmissions. The performance of MaxRate increases when the traffic
load grows beyond 440 frames per second.  It, however, causes the
starvation problem as we have shown in Fig.~\ref{fig:upaccess}.
Overall, our design produces a much higher total throughput as
compared to the others, while ensuring fair access.

\begin{figure}[t]
\centering
\epsfig{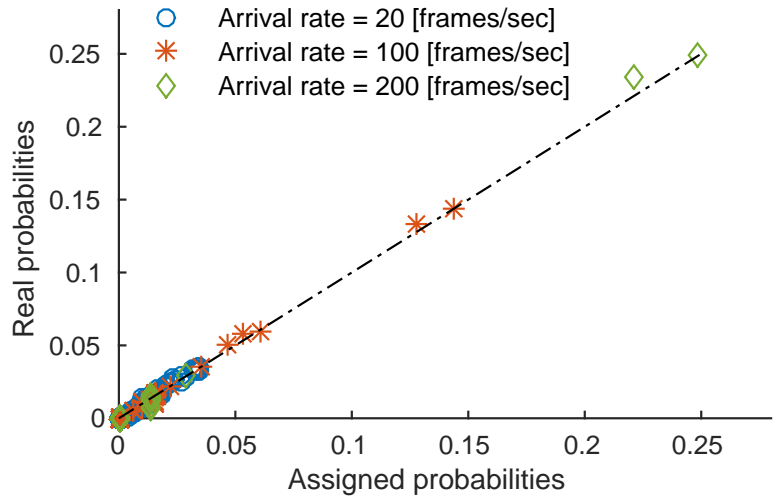}
\caption{Comparison between the assigned probabilities and the contention results}
\label{fig:prob_cdf}
\end{figure}

\begin{figure}[t]
\centering
\epsfig{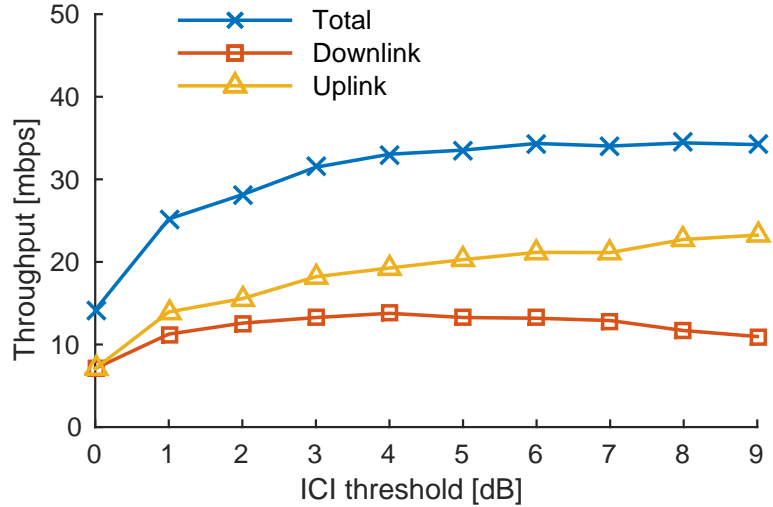}
	\caption{Impact of the ICI threshold}
\label{fig:sinr_threshlod}
\end{figure}

\subsection{Performance of Probabilistic-based Contention}
We now check whether our probabilistic contention design can realize
the theoretical assigned access probability in practice.
Fig.~\ref{fig:prob_cdf} compares the real access probability measured
in the simulations with the assigned probability solved from the
model, when the traffic arrival rate varies from 20 to 200 frames per
second. The results show that the real probabilities are fairly close
to the assigned ones. It confirms that our probability-based
contention can effectively prioritize clients {\em in a distributed
way} according to the assigned probability, no matter the network is
congested or not.  With this distributed probabilistic-based
contention, the AP does not need to explicitly schedule all the
transmissions in a time-division manner.  Our design hence is more
practical to be implemented, and maintains the advantage of
distributed access in the traditional 802.11.


\subsection{Impact of the ICI threshold}
We next examine the impact of the ICI threshold $\delta$ on the
performance of our probabilistic-based MAC protocol.
Fig.~\ref{fig:sinr_threshlod} plots the average downlink throughput,
uplink throughput and total throughput, when the ICI threshold
$\delta$ is set from 0 to 9 dB.  When ICI threshold is 0 dB, we should
ensure that a downlink client cannot hear any interference from the
concurrent uplink at all.  It means that our probability assignment
algorithm can only give hidden nodes full-duplex transmission
opportunities. All the other nodes should fall back to the half-duplex
mode. In this case, the full-duplex opportunities could be reduced
significantly. As the threshold increases, a downlink client can
tolerate an increasing amount of inter-client interference, and,
hence, can allow more full-duplex transmission opportunities.  This
implies that clients have more opportunities to send uplink traffic,
as a result achieving a higher uplink throughput and also a higher
total throughput.  Increasing the threshold, however, introduces a
higher interference to downlink clients, and, hence, slightly degrades
the downlink throughput.  Overall, the total throughput is quite
stable when the ICI threshold exceeds 5 dB. Therefore, considering the
trade-off between performance and fairness, we choose 5 dB as our
default threshold.

\begin{figure}[t]
\centering
{
	\epsfig{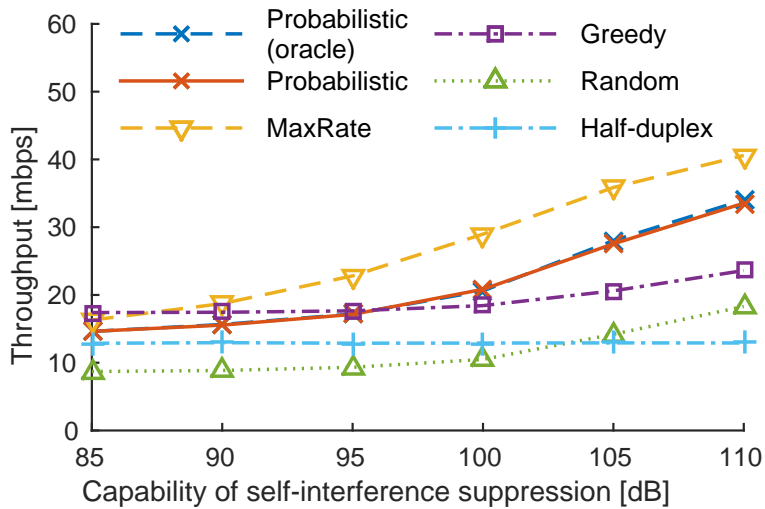}
\caption{Impact of self-interference suppression capability}
\label{fig:sic}
}
\end{figure}

\subsection{Impact of Self-Interference Suppression Capability}
We finally check the impact of self-interference suppression
capability on the throughput gain.  Fig.~\ref{fig:sic} shows that, in
general, the full-duplex gain increases when the AP is able to
suppress more self-interference. The performance gap between the
comparison schemes also increases when the AP has a better suppression
capability because an efficient pairing algorithm can identify
suitable full-duplex pairs from more full-duplex opportunities. Even
with only 85dB interference suppression, which has been verified
feasible in~\cite{ashu}, our design can produce an average throughput
gain of 13\% over half-duplex.  With more advanced full-duplex radio
designs, which can cancel up to 110 dB of
self-interference~\cite{sachin-sigcomm}, the average throughput gain
can be 160\%.

\begin{figure}[t]
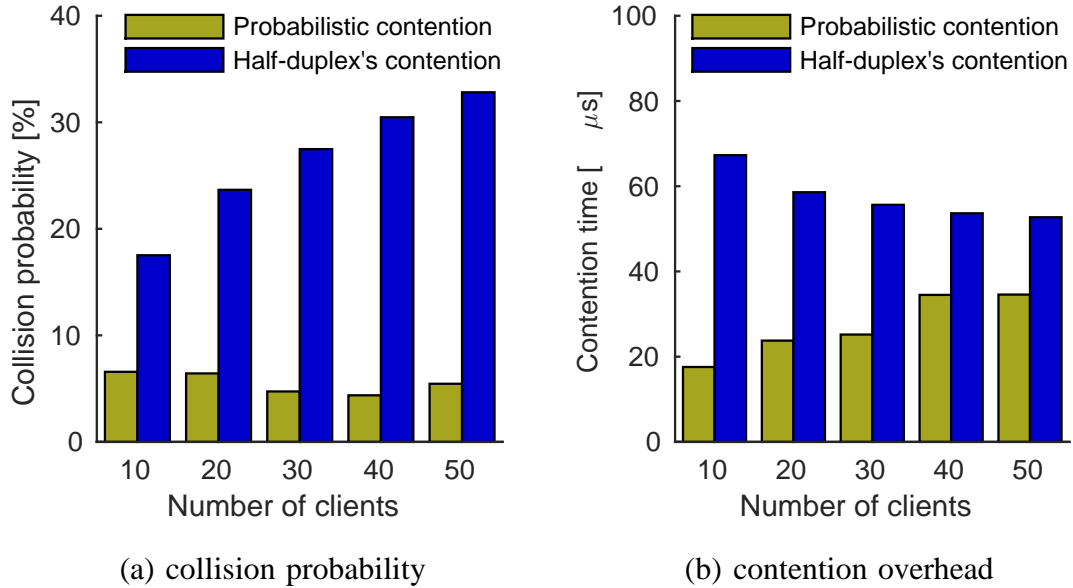

\centering
{
\begin{tabular}{cc}
	\epsfig{width=2.8in,file=collision_probability.eps} &
	\epsfig{width=2.8in,file=contention_time.eps} \\
	(a) collision probability & (b) contention overhead
\end{tabular}
\caption{Protocol effectiveness}
\label{fig:overhead}
}
\end{figure}

\subsection{Protocol Efficiency}\label{sec:result-overhead}
We finally examine efficiency of our protocol design.
Fig.~\ref{fig:overhead}(a) compares the collision probability in our
probabilistic-based contention to that in traditional 802.11
half-duplex contention.  The results show that the proposed MAC not
only delivers the full-duplex gains, but also reduces the collision
probability significantly.  The main reason is that our scheme
leverages frequency-domain contention to reduce the contention
overhead for downlink traffic, and, also, given a downlink
transmission, only part of clients are allowed to contend for
full-duplex uplink transmission.  Not only this, unlike 802.11, which
{\em detects} collisions and then applies exponential random backoff,
we allows clients to {\em avoid} collisions by adjusting its
contention window based on the assigned probabilities.  The clients
that are assigned a higher probability can have a smaller contention
window, and prevent from being collided by other clients. We also show
in Fig.~\ref{fig:overhead}(b) that this probabilistic-based contention
further decreases the required contention time, as a result improving
the resulting effective throughput.
\section{Conclusion} \label{sec:conclusion}
In this paper, we introduced a probabilistic-based MAC protocol for
three-node full-duplex.  We adopted probabilistic client pairings to
best exploit the full-duplex opportunities to deliver the overall
network throughput gain.  Each AP periodically assigns an access
probability to each pair of downlink-uplink clients based on not only
their historical ICI but also their heterogeneous traffic demands. The
APs and clients then use the allocated access probabilities as hints
to contend for full-duplex or half-duplex transmissions using
traditional random backoff in a distributed manner.  To preventing
from affecting the decodability of downlink clients, each uplink
client performs power control to manage inter-client interference on a
per-packet basis.  With the cooperation of uplink clients, the AP can
hence adapt the transmission bit-rate of both the downlink and uplink
packets accordingly and ensure successful decoding in both directions.
We showed via simulations that, by combining probabilistic-based
scheduling and random access, our design efficiently extracts the
full-duplex gains and improves the overall throughput, while
maintaining the distributed nature and fairness provision of 802.11.

\bibliographystyle{IEEEtran}
\bibliography{paper}

\end{document}